\documentclass[aps,prb,preprint,groupedaddress,showpacs]{revtex4}
\usepackage{graphicx}
\usepackage{amsmath}
\usepackage{mathrsfs}

\begin{document}
\title{Dynamics of diluted magnetic semiconductors from atomistic spin dynamics simulations: Mn doped GaAs as a case study}

\author{J. Hellsvik}
\email[]{johan.hellsvik@fysik.uu.se}
\author{B. Skubic}
\author{L. Nordstr\"om} 
\author{B. Sanyal}
\author{O. Eriksson}
\affiliation{Department of Physics and Materials Science, Uppsala University, Box 530, SE-751 21 Uppsala, Sweden}
\author{P. Nordblad}
\author{P. Svedlindh}
\affiliation{Department of Engineering Sciences, Uppsala University, Box 534, SE-751 21 Uppsala, Sweden}
\date{\today}

\begin{abstract}
The dynamical behavior of the magnetism of diluted magnetic semiconductors (DMS) has been investigated by means of atomistic spin dynamics simulations. The conclusions drawn from the study are argued to be general for DMS systems in the low concentration limit, although all simulations are done for 5~\% Mn-doped GaAs with various concentrations of As antisite defects. The magnetization curve, $M(T)$, and the Curie temperature $T_C$ have been calculated, and are found to be in good correspondence to results from Monte Carlo simulations and experiments. Furthermore, equilibrium and non-equilibrium behavior of the magnetic pair correlation function have been extracted. The dynamics of DMS systems reveals a substantial short ranged magnetic order even at temperatures at or above the ordering temperature, with a non-vanishing pair correlation function extending up to several atomic shells. For the high As antisite concentrations the simulations show a short ranged anti-ferromagnetic coupling, and a weakened long ranged ferromagnetic coupling. For sufficiently large concentrations we do not observe any long ranged ferromagnetic correlation. A typical dynamical response shows that starting from a random orientation of moments, the spin-correlation develops very fast ($\sim$ 1ps) extending up to 15 atomic shells. Above $\sim$ 10 ps in the simulations, the pair correlation is observed to extend over some 40 atomic shells. The autocorrelation function has been calculated and compared with ferromagnets like bcc Fe and spin-glass materials. We find no evidence in our simulations for a spin-glass behaviour, for any concentration of As antisites. Instead the magnetic response is better described as slow dynamics, at least when compared to that of a regular ferromagnet like bcc Fe.
\end{abstract}


\maketitle

\section{\label{intro}Introduction}
The interest in dilute magnetic semiconductors (DMS) has been enormous during the last decade.\cite{jungwirth,macdonald,ohno,GaN,dietlnew,ZnO-2,edmonds,dassarma,dietl_science,zungersemicond,akai,pavel,dederichs, Kudrnovsky_2004, Zunger_2004, larsprl,satoprb,larsprb,ky} Numerous theoretical and experimental investigations focusing on their properties have been performed, motivated not least by the quest for a DMS system with ferromagnetic properties at and above room temperature. One of the more frequently studied materials is the III-V semiconductor gallium arsenide, doped with manganese (Mn-GaAs), and it now stands clear that an ordering temperature of 170 K\cite{edmonds} to 180 K \cite{Olejnik08} can be obtained. Among the achievements on the theoretical side is the success of calculations based on density functional theory, where quite generally for DMS systems one should mention the following: first principles calculations result in 'reasonable' electronic band structures\cite{akai} as compared to experiments, the correct values of the magnetic moments are obtained\cite{akai,pavel,dederichs}, calculations of interatomic exchange interactions\cite{Kudrnovsky_2004,dederichs,Zunger_2004} result in ordering temperatures in agreement with observations, and finally the recognition of percolation\cite{larsprl,satoprb,larsprb} as an important mechanism to govern the presence or absence of long ranged magnetic ordering, so that mean field theories should be used with great caution.

Whereas most authors have concentrated on the static properties in thermal equilibrium, the scope of this Article is atomistic simulations of the dynamic properties of diluted magnetic semiconductors. We have chosen to use Mn doped GaAs, with and without defects such as As antisites, as a representative of this group of materials and in our manuscript we argue that our findings apply to DMS in a quite general way. The main reason behind this is that both III-V and II-VI based DMS have very similar behavior as regards the exchange interaction, with an exponentially decaying behavior with distance between magnetic impurities.\cite{Kudrnovsky_2004} In addition, for both sets of systems a varying degree of ferromagnetic and antiferromagnetic interactions can be obtained, depending on vacancy concentration of the host lattice. For instance, Mn doped ZnO has antiferromagnetic interactions which can be turned to ferromagnetic interactions in the presence of Zn vacancies.\cite{ZnO-2} In a similar way, it is known that Mn doped GaAs has primarily ferromagnetic interactions between the Mn atoms in the absence of defects, but an increasing amount of As antisites and Mn interstitials introduce antiferromagnetic interactions between the Mn atoms.\cite{larsprl} 
The main difference between the III-V and II-VI materials is that the solubility of transition metals is larger for the latter, but in the low concentration limit (2-10~\% of transition metals doped into the host lattice) these two sets of materials should have a similar magnetic behavior and dynamic response.

Many DMS materials are in experiments analyzed as changing from a ferromagnetic behavior to a regime where spin-glass behavior is thought to occur. The strongest evidence for spin-glass properties can be found among the II-VI DMS systems, like $\mbox{Zn}_{1-x}\mbox{Mn}_{x}\mbox{Te}$ \cite{shan1998} and $\mbox{Zn}_{1-x}\mbox{Mn}_{x}\mbox{In}_{2}\mbox{Te}_{4}$ \cite{goya2001,peka2007}. Among the III-V DMS materials spin-glass properties have been suggested for $\mbox{GaMnN}$ \cite{dhar2003} and Te doped $\mbox{GaMnAs}$\cite{yuldashev2004}.
Often this transition is found as a function of concentration of magnetic dopants, but the concentration of other defects have also been analyzed to drive a spin-glass behavior. A good example of this is the system of interest here, Mn doped GaAs, where defects like As antisites, have been argued to produce a spin-glass like behavior\cite{katayama2003}. Spin-dynamics simulations, as presented here, are a natural way to analyse this behaviour and part of this study is devoted to analysing spin-glass behaviour in Mn doped GaAs.

In the present paper we give a full account of a spin-dynamics simulation of a DMS material, including an analysis of short versus long ranged correlation and typical time scales of the magnetic response. We have considered Mn$_x$Ga$_{1-x}$As with a Mn concentration of $x=5$~\% and a varying degree $y=0.00-2.00$~\% of As antisite concentration (As atoms which are positioned on the Ga sub-lattice). As this paper reports atomistic spin-dynamics simulations of DMS materials we note that one previous spin-dynamics simulation has been published in Ref.~\onlinecite{jxlu2005}. 

The rest of the paper is organised as follows. In Section~\ref{sect:meth} we present the details of our method, in Section~\ref{sect:res} our results and our conclusions are finally given in Section~\ref{sect:conc}.

\section{\label{sect:meth}Method}
In studies of static properties for diluted magnetic systems, a two step procedure is often used, combining density functional theory (DFT) and Monte Carlo (MC) techniques. We have here also used a two step procedure, but instead of a MC simulation as the second step, the time evolution of the spin dynamics has been explicitly simulated by treating the equation of motions for the atomic magnetic moments. We use atomistic spin dynamics (ASD) as derived in Refs.~\onlinecite{antr1995,antr1996}. The inclusion of temperature in the simulations is treated with Langevin dynamics.\cite{garc1998} A thorough description of the details of our method is given in Ref.~\onlinecite{asd}. The central entity of the technique is the microscopic equations of motion for the atomic moments, $\mathbf{m}_i$, in an effective field, $\mathbf{B}_{i}$, which is expressed as follows,
\begin{equation}
\frac{d\mathbf{m}_i}{dt}=-\gamma \mathbf{m}_i \times [\mathbf{B}_{i}+\mathbf{b}_{i}(t)]-\gamma \frac{\alpha}{m} \mathbf{m}_i \times (\mathbf{m}_i \times [\mathbf{B}_{i}+\mathbf{b}_{i}(t)]).
\label{eq:sllg}
\end{equation}
In this expression $\gamma$ is the electron gyromagnetic ratio and $\mathbf{b}_{i}(t)$ is a stochastic magnetic field with a Gaussian distribution. The magnitude of that field is related to the damping parameter, $\alpha$, in order for the system to eventually reach thermal equilibrium. We have used one heat bath and the time step for solving the differential equations was $\Delta t=0.01$ femtoseconds.

\begin{figure}
\includegraphics[width=0.4\textwidth]{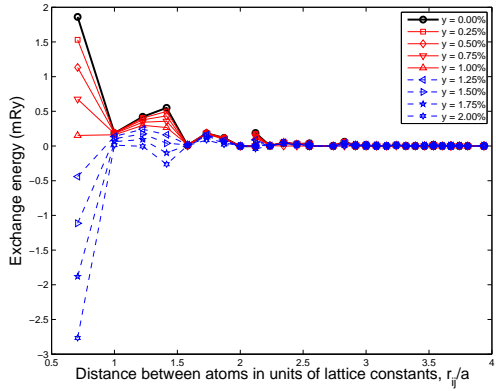}
\caption{\label{fig:exchpar1}(Color online) Calculated exchange parameters for 5~\% Mn doped in GaAs, over 39 atomic shells of the Ga sublattice, with $y=0.00-2.00$~\% the concentration of As antisites. The 39 shells have 29 unique distances, e.g. (0.5 0.5 2.0) and (0.0 1.5 1.5) both have distance $\approx 2.12a$. Data redrawn after Ref.~\onlinecite{Kudrnovsky_2004}.}
\end{figure}

The effective field, $\mathbf{B}_i$, on an atom $\it i$, is calculated from 
\begin{equation}
\mathbf{B}_i=-\frac{\partial \mathscr{H}}{\partial \mathbf{m}_i},
\label{eq:heisenberg}
\end{equation}
where only the part of the Hamiltonian, $\mathscr{H}$, which represents interatomic exchange interactions, $\mathscr{H}_\mathrm{ex}$, is considered in the present simulations. For this we use the classical Heisenberg Hamiltonian, 
\begin{equation}
\mathscr{H}_\mathrm{ex}=-\frac{1}{2}\sum_{i\neq j}J_{ij}\mathbf{m}_i\cdot\mathbf{m}_j,
\end{equation}
where $i$ and $j$ are atomic indices and $J_{ij}$ is the strength of the exchange interaction, which is calculated from first principles theory. The exchange parameters $J_{ij}$ used here were calculated by Kudrnovsk\'y \textit{et al.} from first principles theory as described in Ref.~\onlinecite{Kudrnovsky_2004}. The exchange interactions are strongly direction dependent and have been calculated up to and including the 39:th shell, which were all included in the present spin dynamics simulation (see Fig.~\ref{fig:exchpar1}). Note in Fig.~\ref{fig:exchpar1} that we have plotted the exchange parameters as a function of distance, not angle. As shown in Ref.~\onlinecite{Kudrnovsky_2004} there is a strong angular anisotropy of these exchange interactions. The calculated Mn projected magnetic moment depends on the As antisite concentration,\cite{pavel} and we have used the value of Ref.~\onlinecite{Kudrnovsky_2004}, which is $\sim \mu=4$ $\mu_B/$atom. In principle the Mn projected moment is a little larger than 4, weakly increasing with As antisite concentration. Hence we consider As antisites to influence primarily the exchange interactions, not the size of the magnetic moments, a fact which is consistent with e.g. Refs.~\onlinecite{pavel} and \onlinecite{Kudrnovsky_2004}. Temperature effects of the size of the magnetic moment and strength of the exchange interactions, via changes of the electronic structure, have not been included in this work. 
The pair correlation function for equal times is defined as 
\begin{eqnarray}
G_{ij}(t) & = & \langle \mathbf{m}_i(t)\cdot \mathbf{m}_j(t)\rangle\\
& = & \langle \mathbf{m}(\mathbf{r}_i,t)\cdot \mathbf{m}(\mathbf{r}_j,t)\rangle\nonumber
\end{eqnarray}
where $\langle ... \rangle$ represents an average over all atoms in the simulation box. This quantity will be utilized to illustrate the dynamics as the system approaches thermal equilibrium.   

\section{\label{sect:res}Results}
\subsection{\label{subsect:magnetisation} Simulation of the Magnetisation}
\begin{figure}
\includegraphics[width=0.4\textwidth]{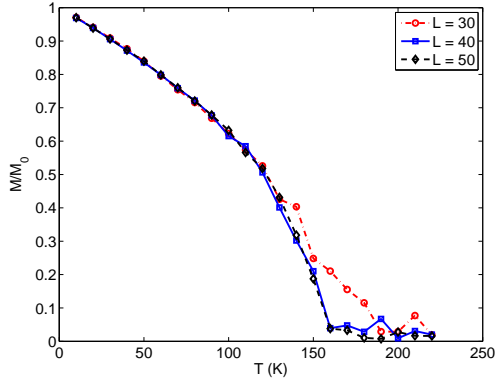}
\caption{\label{fig:magT3}(Color online) Magnetization (normalized to $M_0$ where $M_0$ is the saturation magnetization at $T=0$~K), $M(T)$, versus temperature, $T$, for Mn$_x$Ga$_{1-x}$As with no As antisites, $y=0.00$~\%. $L=30,40,50$ corresponds to different sizes of the supercell.}
\end{figure}

To establish that a sufficiently large supercell was used, calculations were made for systems with size $L\times L\times L$ conventional fcc unit cells, with $L=30,40,50$. Periodic boundary conditions were imposed. With a Mn doping concentration of 5~\%, this corresponds to $5400,12800$ or $25000$ magnetic atoms, respectively. The magnetization versus temperature curve, where the magnetization is the average of the magnetic moments in the supercell, is displayed in Fig.~\ref{fig:magT3}. It is noteworthy that convergence in the magnetization curve is achieved at $L=40$ and for this reason all subsequent calculations were made with this or a higher value of $L$. We also note that Fig.~\ref{fig:magT3} suggests an ordering temperature in the vicinity of 160 K, which is close to the experimental value.\cite{edmonds}

\begin{figure}
\includegraphics[width=0.4\textwidth]{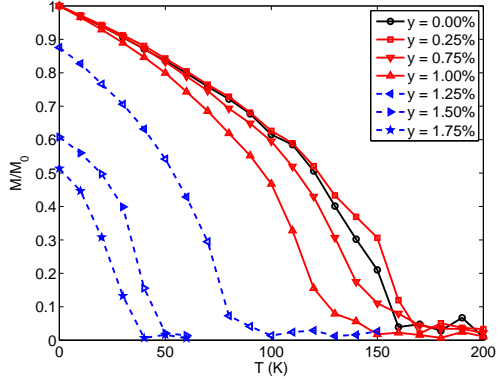}
\caption{\label{fig:magAT7}(Color online) Magnetization $M(T)$ (normalized to the saturation magnetization $M_0$ at $T=0$~K) for Mn$_x$Ga$_{1-x}$As with As antisites. The antisite concentration ranges from $y=0.00$~\% to $y=2.00$~\%.}
\end{figure}

As the concentration of As antisites is increased, the interatomic exchange couplings change from being purely ferromagnetic to having also an antiferromagnetic component. This has been shown to reduce the ordering temperature, both in experiment\cite{edmonds} and in theory.\cite{pavel,larsprl,satoprb} The temperature dependence of the magnetization from the ASD simulations is plotted in Fig.~\ref{fig:magAT7}, for different values of the As antisite concentration $y$. We note that the general behavior of the data in Fig.~\ref{fig:magAT7} and that of Refs.~\onlinecite{larsprl,larsprb} is the same. For a given set of pair exchange parameters ASD and MC simulations should indeed give the same result for the $M(T)$ function and $T_\mathrm{C}$. This follows as statistical mechanical considerations for thermal equilibrium require the energy of the magnetic degrees of freedom to be distributed according to a Boltzmann function \cite{asd} in both methods. In Table.~\ref{tab:tctab} our estimated values of the ordering temperature are seen to match fairly well with those calculated in Ref.~\onlinecite{larsprb}. The MC values for $T_C$  were calculated using a fourth order cumulant method. Our ASD values for $T_C$ were estimated from the points where the $M(T)$ curves approaches zero. That circumstance might explain that our values are slightly higher than those in Ref.~\onlinecite{larsprb}.
\begin{table}
\caption{\label{tab:tctab}Critical temperatures, $T_C$ (K), for different As antisite concentrations, $y$, given in percent. The Mn concentration is 5~\%. MC stand for Monte Carlo data of Ref.~\onlinecite{larsprl} and ASD the current values from spin dynamic simulations.}
\begin{ruledtabular}
\begin{tabular}{|l|c|c|}
As concentration& MC & ASD \\
\hline
y=0.00\% & 137 & 160 \\
y=0.25\% &  -  & 160 \\
y=0.75\% &  -  & 135 \\
y=1.00\% & 92  & 115 \\
y=1.25\% & 55  & 75 \\
y=1.50\% & 26  & 38 \\
\end{tabular}
\end{ruledtabular}
\end{table}

\subsection{\label{subsect:dynamics} Dynamical correlation}
\begin{figure}
\includegraphics[width=0.4\textwidth]{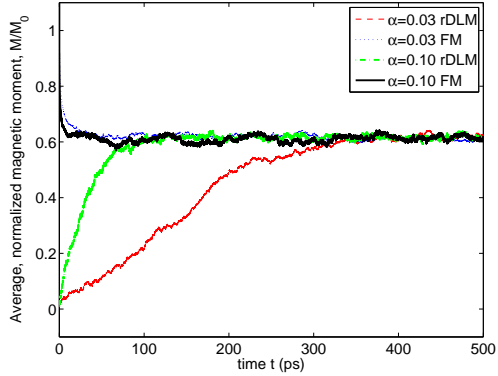}
\caption{\label{fig:avphases}(Color online) Time evolution of the average, normalized magnetization starting from ferromagnetic (blue) respective random (red) spin configurations for $L=40$, with As Antisite concentration $y=0.25$~\% at temperature $T=100$ K and with a damping parameter of 0.03. Similar simulations but with a damping parameter of 0.1 are shown in green and black for the random and ferromagnetic configurations, respectively.}
\end{figure}

The advantage of the spin dynamics approach over Monte Carlo simulations is that the true dynamics of the magnetism may be captured. As an example of this we show the time evolution of the macroscopic magnetic moment of our simulation box as a function of temperature, starting from a ferromagnetic (FM) or a completely random disordered local spin configuration (rDLM). In the latter configuration, each spin is with uniform probability assigned a direction on the unit sphere, independently of the other spins. In Fig.~\ref{fig:avphases} we show data for an As antisite concentration of $y=0.25~\%$ at temperature $T=100$ K. It should also be noted that the relaxation behavior is to some extent dependent on the damping parameter. In most of our simulations presented here we have used the value $\alpha=0.10$, but in Fig.~\ref{fig:avphases} we also show data for $\alpha=0.03$. Although in general smaller differences between results from the two damping parameters can be detected, we note that the main conclusions of our study are not influenced by this choice. A value of $\alpha=0.10$ is slightly higher than recent values for the damping in Mn$_x$Ga$_{1-x}$As, primarily obtained from experiments\cite{xliu2003} but also from theoretical estimates.\cite{sino2004}

Figure~\ref{fig:avphases} shows that the ferromagnetic starting configuration relaxes faster to the equilibrium value, which is natural since the equilibrium configuration is closer to the ferromagnetic configuration than the rDLM configuration. The relaxation time is also different since the mechanisms to reach equilibrium are different. Starting from the ferromagnetic configuration the approach to equilibrium is primarily controlled by spin precession (magnon excitations), while starting from the rDLM configuration the approach to equilibrium involves both spin precession and spin flip processes. Since the relaxation in the two cases is different it is natural that the relaxation rates are different. Figure~\ref{fig:avphases} also shows that the typical time-scale for the relaxation is $\sim 5-10$~ps for the ferromagnetic configuration and $\sim 100$~ps for the rDLM configuration at a $T/T_C$ ratio $100/160$, using a damping $\alpha=0.1$ and cell size $L=40$. Graphs are also shown for $\alpha=0.03$ and it may be seen that in this case the time to reach equilibrium is approximately 250-300 ps. Hence when starting from a rDLM configuration the time to reach equilibrium seems to scale approximately linearly with the value of the damping parameter.

A comparison can be made concerning the typical relaxation times for the system of interest here, Mn doped GaAs, and a typical ferromagnet, bcc Fe. When doing such a comparison it is important to have similar values of the simulation cell, the damping parameter and the T/T$_C$, for the two systems, and we have indeed done this. The comparison was made at T/T$_C$= 0.62 (data for Fe not shown) and shows that the dynamics of Mn doped GaAs is slower with approximately 70 \% than that of bcc Fe. To some degree this may be explained by the weaker local exchange field (${\bf B}_i$ in Eq. 1) of the Mn doped system, due to that it is a diluted system.

\begin{figure}
\includegraphics[width=0.4\textwidth]{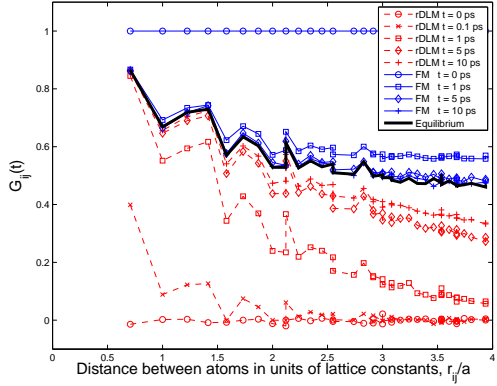}
\caption{\label{fig:corrt4A002T100}(Color online) Time evolution of the pair correlation function $G_{ij}(t)$ starting from ferromagnetic (blue) respective random (red) configurations and with $L=40$ in the simulation. Values are obtained for $y=0.25$~\% and $T=100$ K. The equilibrium pair correlation is shown in black.}
\end{figure}

\begin{figure}
\includegraphics[width=0.4\textwidth]{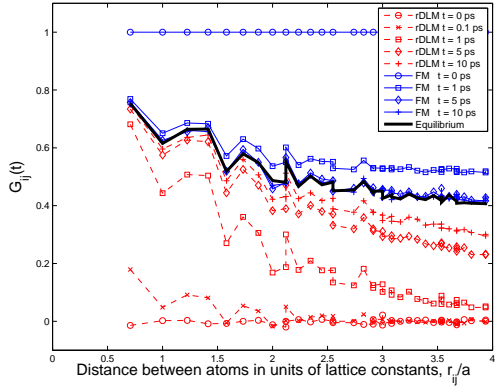}
\caption{\label{fig:corrt4A007T100}Same as Fig.~\ref{fig:corrt4A002T100} but for $y=0.75$~\%.}
\end{figure}

\begin{figure}
\includegraphics[width=0.4\textwidth]{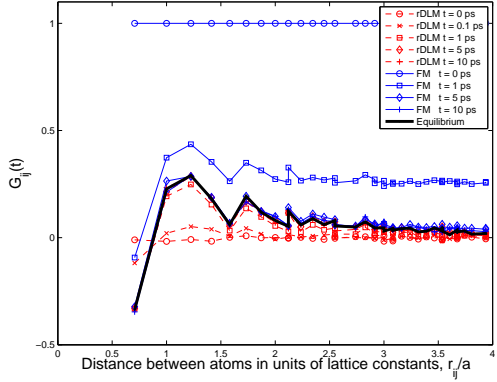}
\caption{\label{fig:corrt4A012T100}Same as Fig.~\ref{fig:corrt4A002T100} but for $y=1.25$~\%.}
\end{figure}

The pair correlation function, $G_{ij}(t)$, depends not only on the distance between Mn atoms but also, as an immediate consequence of the directional dependence of the exchange interaction, on the direction in the lattice in which $G_{ij}(t)$ is calculated. In Figs.~\ref{fig:corrt4A002T100}-\ref{fig:corrt4A012T100} the pair correlation function is plotted at different times as a function of distance between atoms. In these figures the deviation from a smooth decreasing behavior reflects indeed the angular anisotropy of the exchange interaction. Note that in these simulations we have used a temperature of 100 K and an As antisite concentration of 0.25~\%, 0.75~\% and 1.25~\%. For the As 1.25~\% antisite concentration the ordering temperature is below 100 K, but for the other As antisite concentrations the ordering temperature is above 100 K. The simulations were started both from a rDLM and a FM configuration.
A few general remarks can be made concerning the data in Figs.~\ref{fig:corrt4A002T100}-\ref{fig:corrt4A012T100}. For instance, nearest neighbor (NN) spins reach their equilibrium values faster (typically after 1 ps), whereas with increasing distance the dynamics of the correlation slows down (5-10 ps). It is also clear that for Figs.~\ref{fig:corrt4A002T100} and ~\ref{fig:corrt4A007T100} the FM starting configuration seems to reach equilibrium faster than the rDLM configuration, whereas for Fig.~\ref{fig:corrt4A012T100} the rDLM configuration reaches equilibrium faster than the FM configuration. It is natural that the systems shown in the former two figures reach equilibrium from the ferromagnetic configuration faster, since these systems are below the ordering temperature, where a ferromagnetic starting point should be closer to the equilibrium configuration. We also note that for all systems there is a very strong short ranged order, involving several neighboring atomic shells, which is present at all temperatures, even up to and above the ordering temperature (cf. in Fig.~\ref{fig:corrt4A012T100}).

Turning to the details of the curves we note that e.g. in Figs.~\ref{fig:corrt4A002T100} and ~\ref{fig:corrt4A007T100} the NN relaxation time is close to 1 ps, both for the rDLM and FM starting configurations. The long ranged (39 shell distance) relaxation is $\sim$5 ps for the FM starting configuration and $\ge$ 10 ps for the rDLM configuration. For the system which is above the ordering temperature a slightly different behavior is observed. In Fig.~\ref{fig:corrt4A012T100} the rDLM configuration reach equilibrium after 1 ps, for the NN and longer ranged distances. The FM configuration is here relaxing slower, requiring 5 ps. To some degree, the dynamics above the ordering temperature is similar to that below the ordering temperature.

\begin{figure}
\includegraphics[angle=0,width=0.4\textwidth]{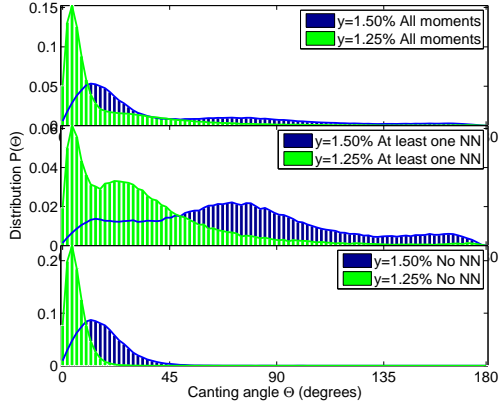}
\caption{\label{fig:cantangleABC}(Color online) Canting for $y=1.25\%$ (green) respective $y=1.50\%$ (blue) at 0 K. The upper panel shows the distribution of angles for all moments in the simulation cell. The middle panel shows the distribution for moments with at least one NN. Data for the disjoint set of moments with no NN:s are shown in the lower panel.}
\end{figure}

The data in Figs.~\ref{fig:corrt4A002T100}-\ref{fig:corrt4A012T100} show that as the As antisite concentration increases, antiferromagnetic, superexchange interaction between the Mn-atoms becomes more dominant. This is especially obvious for the NN interaction. This result is consistent with the analysis presented in Ref.~\onlinecite{Bouzerar07} where a non-collinear coupling between NN Mn atoms was found when competing ferromagnetic and antiferromagnetic interactions play role. We have plotted the distribution $P(\theta)$ of angles between the local moments and the average magnetization in the upper panel of Fig.~\ref{fig:cantangleABC} for $y=1.25\%$ and $y=1.50\%$ at $T=0$~K. The distribution of angles sums up to an average magnetization of $M/M_0=0.87$ for $y=1.25\%$ and $M/M_0=0.61$ for $y=1.50\%$ (cf. Fig.~\ref{fig:magAT7}). To analyze $P(\theta)$ the distribution of angles for the disjoint sets of moments with no NN:s respective at least one NN are plotted. In the middle panel of Fig.~\ref{fig:cantangleABC} only the moments with at least one NN have been considered. The antiferromagnetic coupling between NN:s makes sure that a non-negliable fraction of these moments are close to being antiparallel with the direction of the average magnetization. In the lower panel are plotted $P(\theta)$ for local moments that do not have any NN:s. With no strong antiferromagnetic coupling acting on these moments their angles towards the average magnetization is effectively confined to the intervals $0-25^{\circ}$ ($y=1.25\%$) and $0-45^{\circ}$ ($y=1.50\%$).

Our distributions $P(\theta)$ for moments with at least one NN can be compared with Fig. 5 in Ref.~\onlinecite{Bouzerar07}. For antisite concentration $y=1.25\%$ the ASD simulations have a peak in the intervall of $4-6^{\circ}$ where the canted spin model (CSM) instead has a delta peak for $\theta=0^{\circ}$. The mean angle of the respective distribution differ substantially ($\bar{\theta}_{ASD}=41^{\circ}$, $\bar{\theta}_{CSM}\approx 30^{\circ}$ ). For antisite concentration  $y=1.50\%$ the difference in mean angles is small ($\bar{\theta}_{ASD}\approx 82^{\circ}$, $\bar{\theta}_{CSM}\approx 83^{\circ}$ ) although the shape of the distributions are different. The very same exchange couplings\cite{Kudrnovsky_2004} were used for the present ASD simulations and the CSM study in Ref.~\onlinecite{Bouzerar07}. We conclude that the various approximations in the two schemes account for the differences in the histograms of the angles.

In addition, the possibility of non-collinear magnetism in zinc-blende MnAs was discussed in Ref.~\onlinecite{Sanyal06}, which may be important for understanding samples with an inhomogeneous distribution of Mn atoms where Mn rich regions are created. Inside such a region, where the composition locally is close to MnAs, non-collinear magnetic couplings between the Mn atoms are likely to occur.

\begin{figure}
\includegraphics[angle=0,width=0.4\textwidth]{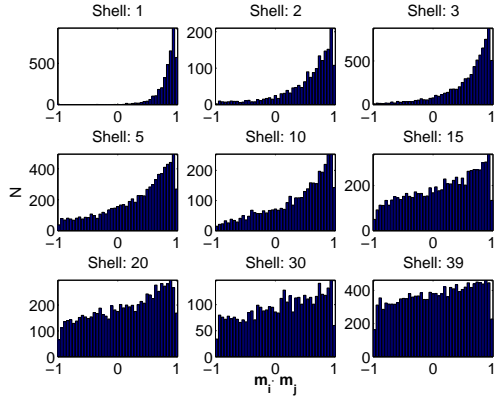}
\caption{\label{fig:corrhist1}(Color online) Histogram of the spin product function $\mathbf{m}_i\cdot \mathbf{m}_j$ starting from random configurations after 1 ps. The simulation box had $L=40$. The temperature is 100 K and the As antisite concentration is 0.25~\%. The value of the spin product is given on the x-axis and the number of atoms with a given spin product is given on the $y$-axis. The pair correlation between different atomic shells is marked in each panel with the shell number in question (e.g. shell 1, shell 2 etc.). The grading of the $y$-axis varies with the number of atoms in the respective shell.}
\end{figure}

\begin{figure}
\includegraphics[angle=0,width=0.4\textwidth]{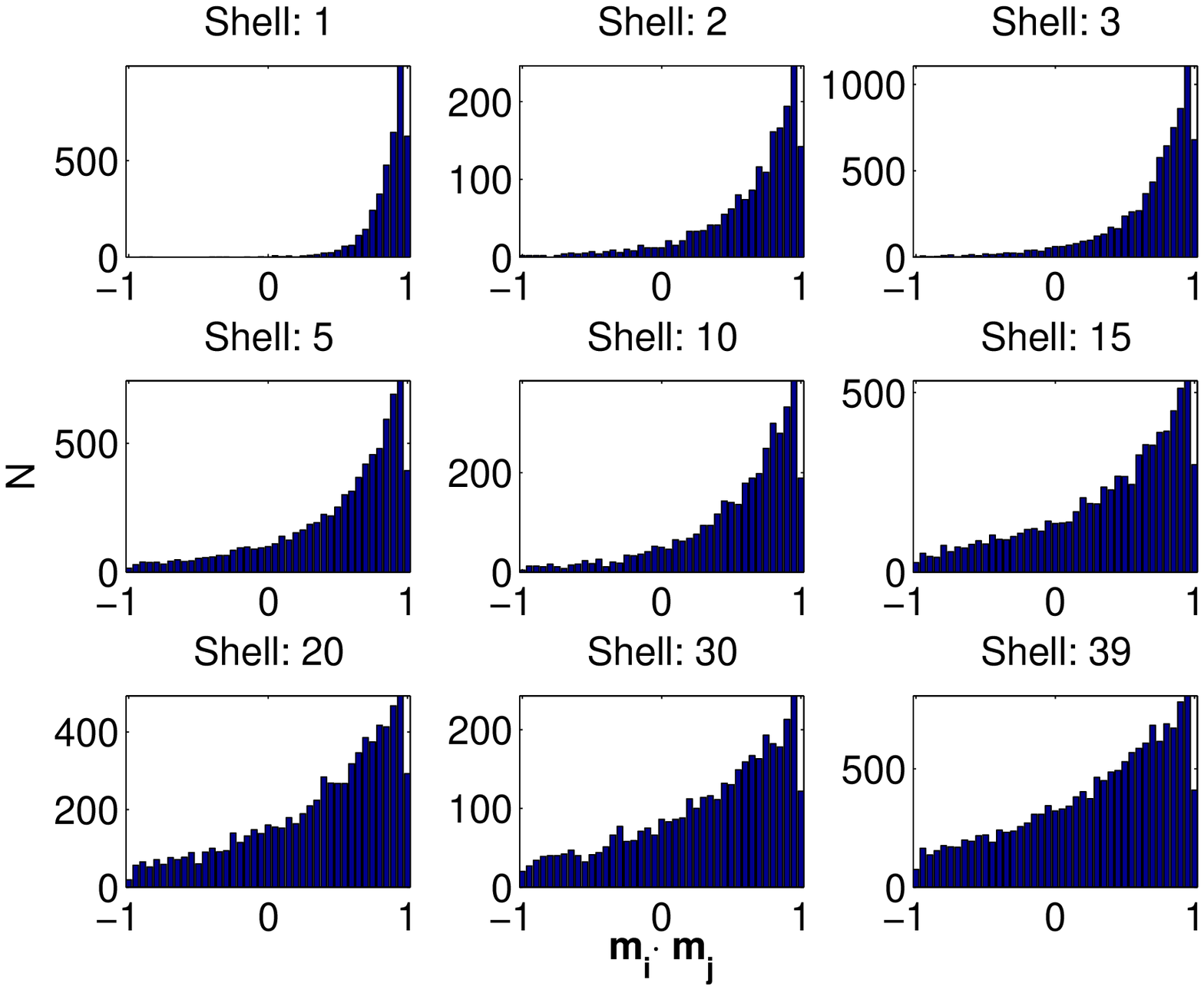}
\caption{\label{fig:corrhist5}(Color online) Same as Fig.~\ref{fig:corrhist1} but after 5 ps.}
\end{figure}

\begin{figure}
\includegraphics[angle=0,width=0.4\textwidth]{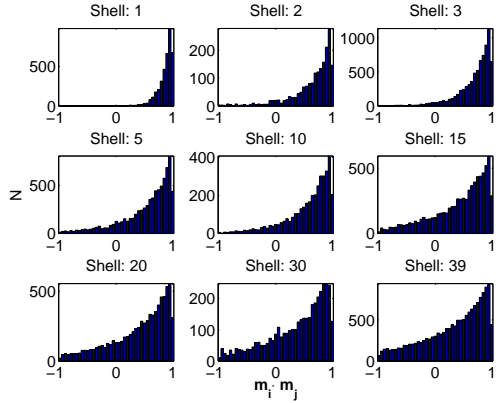}
\caption{\label{fig:corrhist10}(Color online) Same as Fig.~\ref{fig:corrhist1} but after 10 ps.}
\end{figure}

The distribution of the individual scalar products $\mathbf{m}_i\cdot \mathbf{m}_j$ (which we will refer to as the spin-product) for a simulation at 100 K and with a defect concentration of 0.25~\% is shown as histograms in Figs.~\ref{fig:corrhist1},\ref{fig:corrhist5} and \ref{fig:corrhist10} for times $t=1,5$ and $10$ ps, respectively, from a simulation which starts from a completely random configuration of spins. The histograms were calculated as follows. At a given time, $\mathbf{m}_i\cdot \mathbf{m}_j$ was calculated between a given selected Mn atom and all the Mn atoms in the nearest neighboring shell (shell 1), and between the chosen Mn and all Mn atoms in the next nearest neighboring shell (shell 2), etc. This was then repeated over all Mn atoms in the simulation box. In this procedure all spin-products are summed twice and we have corrected for this fact. The number of spin-products with a given value are then shown in histograms Figs.~\ref{fig:corrhist1}-\ref{fig:corrhist10} for various times in the simulation. 

Figures~\ref{fig:corrhist1}-\ref{fig:corrhist10} show that after 1 ps the NN spin-product is equilibrated to its saturation, whereas spin-products between Mn atoms further apart (shell 2-10) have not quite reached equilibrium, but is nevertheless deviating strongly from completely disordered spins (which would show a uniform distribution). An almost  uniform distribution is the general behavior for shell 39. After 5 ps into the simulation (Fig.~\ref{fig:corrhist5}) the spin product between almost all shells have reached their equilibrium distribution and deviate very little from the data shown for 10 ps (Fig.~\ref{fig:corrhist10}). The general behavior of Figs.~\ref{fig:corrhist1}-\ref{fig:corrhist10} is rather similar to a droplet model of magnetic correlations, e.g. as used to analyze spin-glass materials, in that a correlation around a given spin grows outwards in time after the simulation has started\cite{Fisher88,Fisher88-1,Koper88}.

The data in Figs.~\ref{fig:corrt4A002T100}-\ref{fig:corrhist10} suggest that local relaxations cease after some 10-20 ps for a rDLM configuration. These relaxation times are some ten times faster than what Fig.~\ref{fig:avphases} suggests for the same configuration. The reason for this difference is that the data in Figs.~\ref{fig:corrt4A002T100}-\ref{fig:corrhist10} measure time scales of local correlations, whereas the data in Fig.~\ref{fig:avphases} reflects the whole simulation cell. Taken together, the data in Figs.~\ref{fig:avphases}-\ref{fig:corrhist10} suggest that local spin-droplets develop after some 10-20 ps, and that the correlation between such droplets takes 5-10 times longer to develop.


\subsection{\label{subsect:aging}spin-glass analysis, aging and autocorrelation}
The possibility of spin-glass behavior among DMS materials, as discussed in the Introduction, naturally calls for a theoretical analysis, and the spin-dynamics simulations used in this work is a good method for undertaking such an analysis. Hence, from our simulations we have studied aging phenomena, which are typical for spin-glasses, on 5~\% Mn doped GaAs with various concentrations of As antisites. The reason for using the As antisite concentration as a parameter is that one gradually increases the amount of anti-ferromagnetic interactions in the lattice, which starts from purely ferromagnetic interactions, when the As antisite concentrations is zero (see Fig.~\ref{fig:exchpar1}). 
\begin{figure}
\includegraphics[width=0.4\textwidth]{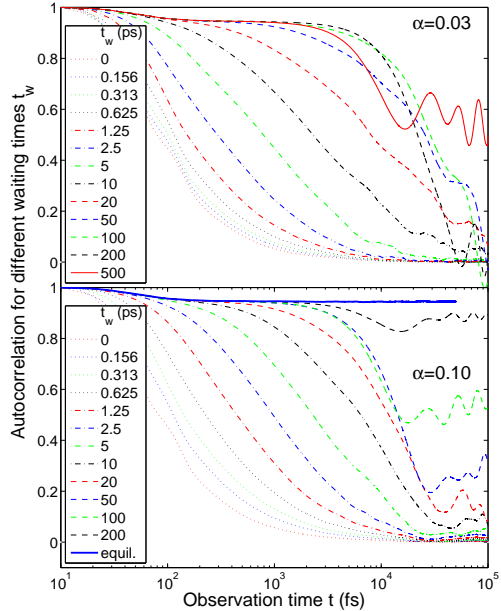}
\caption{\label{fig:acA0025T010AxRx}(Color online) Autocorrelation $C_0(t_w + t,t_w)$ for 5~\% Mn doped GaAs, with 0.25~\% As antisite concentration. Simulations were made at $T=10$~K, and $\alpha=0.03$ (top panel) and $\alpha=0.10$ (bottom panel).}
\end{figure}
\begin{figure}
\includegraphics[width=0.4\textwidth]{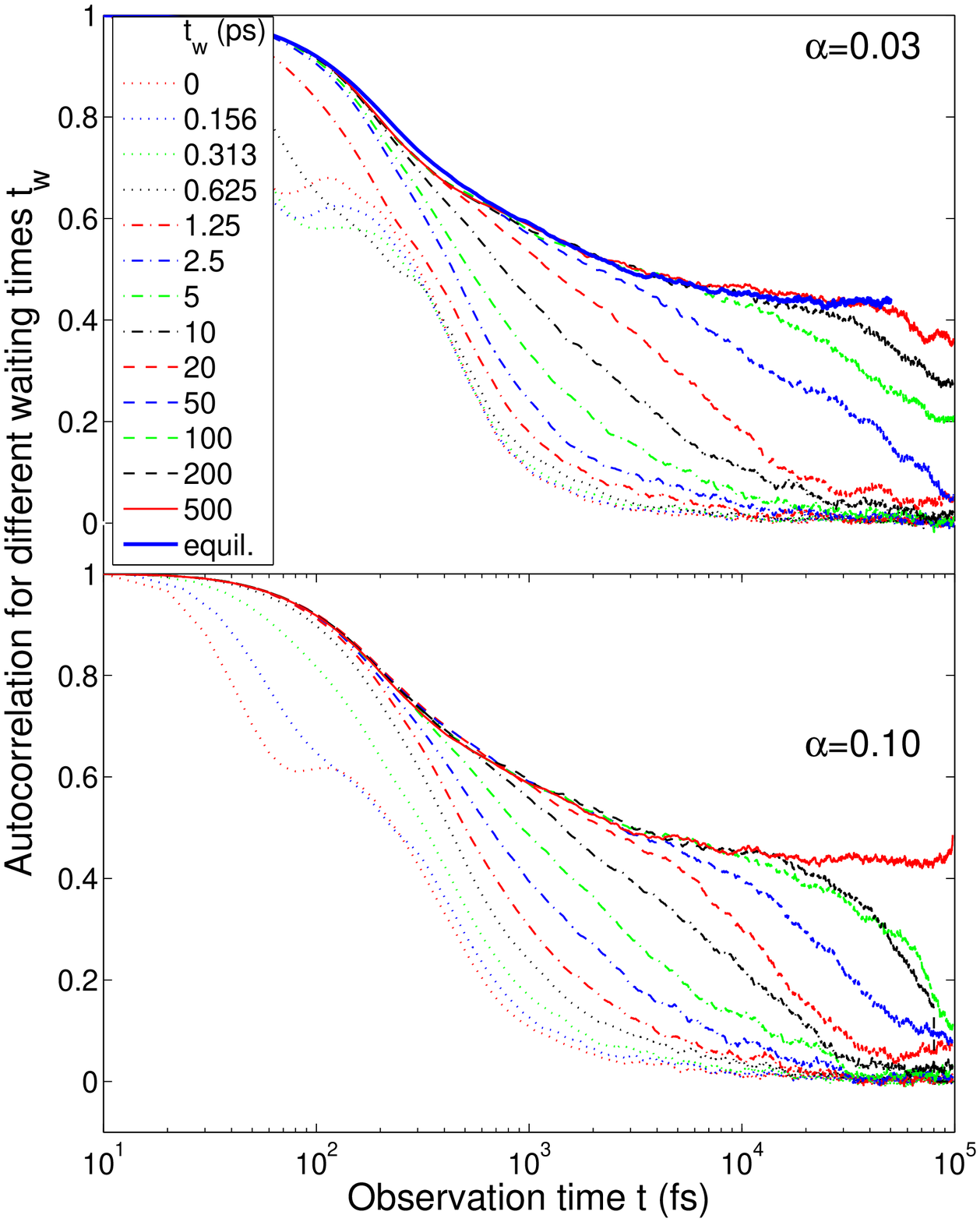}
\caption{\label{fig:acA0175T010A1R1}(Color online) Same as in Fig.~\ref{fig:acA0025T010AxRx} but for 1.75\% As antisites.}
\end{figure}

A slowly relaxing system has physical observables that break time translation invariance before it eventually equilibrates. Two-time observables can sample the characteristics of the approach to equilibrium. The spin autocorrelation function 
\begin{equation}
C_0(t_w + t,t_w) = \langle\mathbf{m}_i(t_w)\cdot\mathbf{m}_i(t_w + t)\rangle
\label{eq:ac}
\end{equation}
is the simplest two-time quantity that has been used to study the aging of spin-glasses \cite{berthier2004}. The autocorrelation is not directly measurable in experiments, where instead the closely related zero (in practice low) field dynamic susceptibility $\mathbf{M}_i(t_w,t)/h$ has been used \cite{nordblad1998}. The calculation of $C_0(t_w + t,t_w)$, was done by starting a simulation from a completely random configuration of Mn moments. A waiting time, $t_w$, was selected and the scalar product between a given moment ${\bf m}_i (t_w)$ at this time and the same moment a time $t$ later (i.e. ${\bf m}_i (t_w+t)$ ) was evaluated. Note that Eq.~\ref{eq:ac} also contains an average over all magnetic atoms of the simulation box.

\begin{figure}
\includegraphics[width=0.4\textwidth]{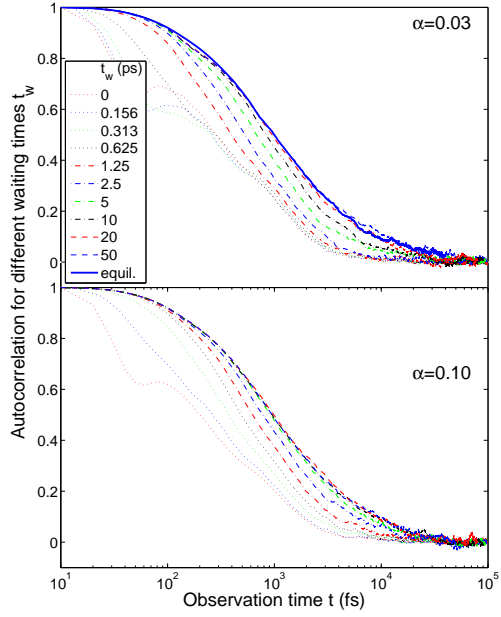}
\caption{\label{fig:acA0200T010A1R1}(Color online) Same as in Fig.~\ref{fig:acA0025T010AxRx} but for 2.00\% As antisites.}
\end{figure}

\begin{figure}
\includegraphics[width=0.4\textwidth]{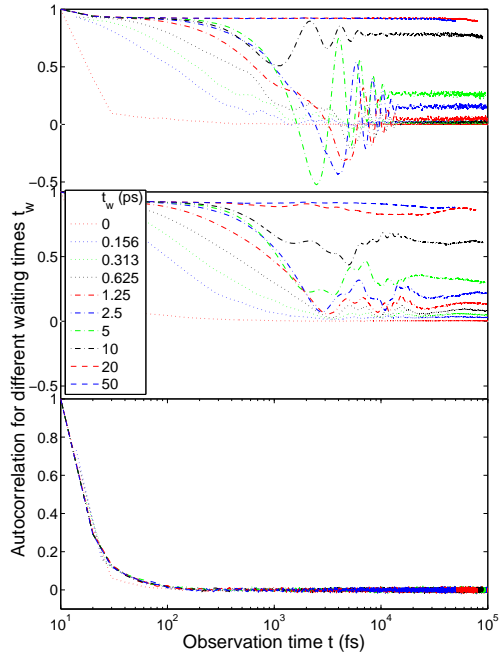}
\caption{\label{fig:acbccFeT100K}(Color online) Autocorrelation $C_0(t_w + t,t_w)$ for bcc Fe at $T=100$~K for a single run (upper panel) and average over 20 runs (middle panel). For comparison also the autocorrelation at $T=1500$~K is shown (lower panel). In the simulations we had $\alpha=0.10$.}
\end{figure}

Before entering the details and the analysis of the autocorrelation functions of Mn doped GaAs, shown in Fig.~\ref{fig:acA0025T010AxRx} (for 0.25~\% As antisites) , Fig.~\ref{fig:acA0175T010A1R1} (for 1.75~\% As antisites) and Fig.~\ref{fig:acA0200T010A1R1} (for 2.00~\% As antisites), we note that data shown for Mn doped GaAs are obtained from averages over up to 20 simulations. In making these averages we used 4 different configurations of Mn atoms, and we used 5 different seeds for generating the heat baths used in the simulations. It was found that making an average over different configurations was more important for obtaining reliable curves to analyse. For comparison, the autocorrelation of a ferromagnetic system (bcc Fe) below and above the ordering temperature is discussed briefly (Fig.~\ref{fig:acbccFeT100K}).

The choice of concentrations of As antisites was made to investigate the dynamics as a function of increasing antiferromagnetic interactions, in samples which have a net ferromagnetic moment (Figs.~\ref{fig:acA0025T010AxRx} and~\ref{fig:acA0175T010A1R1}) and to compare this behavior with a system for which there is no long ranged order (but with finite atomic Mn moments, Fig.~\ref{fig:acA0200T010A1R1}). The data in Figs.~\ref{fig:acA0025T010AxRx},\ref{fig:acA0175T010A1R1} and \ref{fig:acA0200T010A1R1} show that the main characteristics of the autocorrelation functions studied here are not influenced by the choice of damping parameter. The figures also show that for the systems with finite magnetization at $T=10$~K  (Figs.~\ref{fig:acA0025T010AxRx} and \ref{fig:acA0175T010A1R1}) the autocorrelation function approaches a finite value with increasing time, $t$, for long enough waiting times, $t_w$. The data in Figs.~\ref{fig:acA0025T010AxRx} and \ref{fig:acA0175T010A1R1} should be compared to a normal ferromagnetic material, like bcc Fe, and for this reason we show autocorrelation functions for this well-known ferromagnet at T=100 K in Fig.~\ref{fig:acbccFeT100K} (upper panel). Note that for bcc Fe all autocorrelation functions reach a saturation value faster than it does for Mn doped GaAs, but that otherwise the shape of the autocorrelation functions are not too different. 

In section \ref{subsect:dynamics} we discussed how the system relaxes from a rDLM or FM configuration to reach a thermally equilibrated configuration. This discussion was based on the values of the spatial correlation function at different times and will here be extended with the features that can be seen in the autocorrelation curves. For long enough waiting times (curve marked $equil$ in the lower panel of Fig.~\ref{fig:acA0025T010AxRx} and $t_w\ge 10$ ps in Fig.~\ref{fig:acbccFeT100K}), the system has already reached thermal equilibrium. Here the autocorrelation, starting from unity at an observation time $t=0$, quickly settles to a value that equals the value of the magnetization the system posseses at that temperature. The individual spins fluctuate and precess around the direction of the effective magnetic field and their individual autocorrelation functions fluctuate correspondingly. Averaged over the ensemble these fluctuations vanishes and the autocorrelation function, after the initial decay from unity, is constant. For short waiting times ($0\le t_w\le 2.5$~ps in the lower panel of Fig.~\ref{fig:acA0025T010AxRx} and $0\le t_w\le 0.625$~ps in Fig.~\ref{fig:acbccFeT100K}), the autocorrelation quickly drops from unity to essentially zero. 

The more interesting features are to be found for intermediate waiting times ($5\le t_w\le 200$~ps in the lower panel of Fig.~\ref{fig:acA0025T010AxRx} and $1.25\le t_w\le 5$~ps in Fig.~\ref{fig:acbccFeT100K}). The autocorrelation initially decays to a value equal or slightly smaller than the value for the longer waiting times. With increasing observation time the autocorrelation eventually drops from this level and starts a damped oscillation. The value around which the autocorrelation oscillates corresponds to the value the magnetization has reached for a specific waiting time. When comparing the frequencies of the oscillations in Figs.~\ref{fig:acA0025T010AxRx} and ~\ref{fig:acbccFeT100K} (which represent an average over all spins in the simulation cell) with the typical frequencies of individual spin trajectories, we observe that they are of comparable magnitude. The observed oscillations can therefore be explained from the fact that each atomic spin precesses in its local exchange field and that this spin precession is faster than the rate at which the local exchange field reaches its equilibrium orientation. Summing over all spins in the simulation box then gives rise to the oscillatory behaviour in Figs.~\ref{fig:acA0025T010AxRx} and ~\ref{fig:acbccFeT100K}. The damping term in Eq. 1 ensures that the oscillations are damped. If one would make an average over  different configurations of the Mn atoms and different heatbaths, the oscillation diminish in magnitude with the number of simulations. This is illustrated in the middle panel of Fig.~\ref{fig:acbccFeT100K}, where the oscillations are seen to have been strongly suppresed.  

For Mn doped GaAs with 2.00~\% As antisites (Fig.~\ref{fig:acA0200T010A1R1}) the situation is very different compared to the simulations with lower As antisite concentration, since the autocorrelation function approaches zero, for all waiting times. The reason is that the sample for $T=10$~K is paramagnetic. We note however that there are signs of slow dynamics in the simulations, since the different autocorrelation functions do not lie on top of one another, but are spread out over a time interval which covers approximately one decade in time. This is distinctly different from bcc Fe in the paramagnetic phase, at temperatures well above the ordering temperature (see Fig.~\ref{fig:acbccFeT100K} - lower panel), where all autocorrelation functions lie on top of one another, independent on waiting time.
 
For very low temperatures, one can expect a phase transition also for Mn doped GaAs with 2.00~\% As antisites. For a pure spin-glass system, the spin dynamics will exhibit critical slowing down on approaching the transition temperature from above. Below the spin-glass transition temperature, the different autocorrelation functions should not approach zero for any time t, and curves corresponding to different waiting times should never lie on top of one-another for any extended interval of time. Our simulations for $T=2$~K (data not shown) do indeed show a very slow dynamics. For instance, starting from a ferromagnetic configuration or from a rDLM configuration we do not reach equilibrium within the time of the simulation. However, for computer simulations made with limited cell size, the system will eventually exhaust the length scales allowed by the periodic boundary conditions and the autocorrelation function will eventually go to zero. In Ref.~\onlinecite{berthier2004} Berthier \textit{et al.} treats the mechanism of finite size effects in Monte Carlo simulations on a archetypical Heisenberg spin-glass. Problems connected to finite size are also present in our spin dynamics simulations, and we are left to conclude that our data on 2.00~\% at $T=2$~K are not sufficient to allow for any definitive statements on the nature of the phase transition.

It is meaningful to compare the results of Figs.~\ref{fig:acA0025T010AxRx},\ref{fig:acA0175T010A1R1} and \ref{fig:acA0200T010A1R1}, with simulations of typical, well known spin-glass systems, like Mn doped Cu (fcc). We have undertaken simulations on Mn doped Cu (data not shown) and we find that these simulations do indeed show typical spin-glass characteristics, but details of this simulation are outside the scope of the present analysis and will not be discussed further here. Nevertheless, we conclude that our simulation package results in the expected behavior for well established spin-glass systems. The fact that Mn doped GaAs, for several concentrations of As antisites, with varying degree of competing ferromagnetic and antiferromagnetic exchange interactions, does not display spin-glass characteristics suggest that it is incorrect to label this system as a spin-glass.


\subsection{\label{subsect:vis}Visualization}
Scientific visualization techniques enables further insight into spin dynamic simulations. We have produced a set of movies that are accessible on the webpage of Ref.~\onlinecite{asdwww}. These simulations were done without As antisites at a temperature of 100 K (200 K) which is below (above) the ordering temperature. The movies reveal that also above the ordering temperature the spins are correlated on short distances. The difference occurs for longer distances, where for 100 K, the spins are still correlated, but for 200 K they are not.

\section{\label{sect:conc}Conclusions}
In conclusion we have applied atomistic spin dynamics simulations to 5~\% Mn doped in GaAs, with and without As antisite defects. It should be noted that also other types of defects have been found experimentally for this system, e.g. Mn interstitials. Both types of defects introduce antiferromagnetic interactions in the lattice and for simplicity we have here only considered the As antisites. Our results show ordering temperatures and a general behavior of the magnetization curve which are in agreement with Monte Carlo simulations and experiment. 

The dynamical response has also been investigated and we find that the dynamics between NN Mn atoms is, as revealed by the pair correlation function, considerably faster than the more long ranged interaction. In addition we find that short ranged order exists up to and even above the ordering temperature. We also find that starting from random configurations, the pair correlation function, for distances up to $4a$ ($a=$ lattice parameter), reaches equilibrium on a time scale of 10-20 ps. The dynamical response, as revealed by the autocorrelation function, shows that Mn doped GaAs, does not display spin-glass behavior, for moderate ($y=0\%-1.75\%$) concentrations of As antisites. For the highest defect concentration, $y=2.00\%$, the nature of the phase transition (ferromagnetic or spin-glass) of the simulation cell cannot be accurately determined from our data.

The present article focuses on one III-V semiconducting host, GaAs, and one magnetic element, Mn, in the low concentration regime. However, the general behavior of the magnetism of most doped semiconductors is basically similar, with an exchange interaction which decays exponentially with distance due to the presence of a band gap in the system, overlapped with an angular anisotropy of the interatomic exchange, and the possibility to introduce competing ferromagnetic and antiferromagnetic interactions via the defect concentration. For this reason we believe that the main results concerning the dynamics of the presently studied system should be representative for several DMS materials, at least in the diluted limit of magnetic dopants, and that they should be characterized as slow magnets. A notable exception from this conclusion may be the II-VI semiconductors, where the solubility of magnetic atoms can be considerably higher (up to $\sim$ 30~\%) than what we investigate here. For some of these materials\cite{shan1998,goya2001,peka2007} experimental data do indeed show spin-glass behavior. Investigations of such systems are underway.

\begin{acknowledgments}
Calculations have been performed at national computer centres in Sweden (SNIC), UPPMAX and NSC. We thank Josef Kudrnovsk\'y and Lars Bergqvist for providing exchange parameters and Anders Hast for the Visualization ToolKit scripts. We gratefully acknowledge support from the Swedish Research Council (VR) and the Swedish Foundation for Strategic Research (SSF). 
\end{acknowledgments}

\end{document}